\documentclass[11pt]{article}
\usepackage{times}
\usepackage{geometry}
\usepackage[utf8]{inputenc}
\usepackage{enumitem,amssymb}
\usepackage{ragged2e}

\usepackage{fancyhdr}
\usepackage[T1]{fontenc}
\usepackage{epsfig}
\usepackage{wrapfig}
\usepackage{color}

\usepackage{longtable}
\usepackage{caption}
\usepackage[dvipsnames,svgnames,x11names]{xcolor}

\newcommand{\cc}[1]{} %

\definecolor{DarkGreen}{rgb}{0.0, 0.3, 0.0}
\definecolor{purple}{rgb}{0.5, 0.0, 0.5}
\definecolor{red}{rgb}{1, 0.0, 0.0}
\definecolor{green}{rgb}{0, 1.0, 0.0}

\def\3he{$^3{\rm He}$}

\def\lsim{\mathrel{\lower2.5pt\vbox{\lineskip=0pt\baselineskip=0pt
           \hbox{$<$}\hbox{$\sim$}}}}

\def\gsim{\mathrel{\lower2.5pt\vbox{\lineskip=0pt\baselineskip=0pt
           \hbox{$>$}\hbox{$\sim$}}}}

\begin{document}

{\raggedright
\huge
Finding New Debris Discs at Sub-millimetre Wavelengths
\linebreak
\bigskip
\normalsize

\cc{\begin{itemize}
    \item ESO's guidelines and Submission details website: \href{https://next.eso.org/call-for-white-papers/}{https://next.eso.org/call-for-white-papers/}
    \item Submission Deadline: 15 Dec 2025
    \item Page limit: 1 cover page + 3 pages of content
    \item ESO will not make the white papers public but the authors can post them on arXiv.
\end{itemize}}

\cc{As stated on the ESO Expanding Horizons Call for white papers website, they are soliciting white papers from the community (at all career stages) to encourage broad discussions across the community and help identify future challenges.}
\cc{To help coordinate our efforts, the AtLAST team has put together this latex template for internal use, and have decided to make it available more generally for anyone who may find it useful. }
\cc{Some of the blue text guidance in each of the suggested sections is there to help some of our more junior team members understand what to expect to need to include (and at what level). }
\cc{Any suggestions made here reflect how we are intending to answer the call issued by ESO.}
\bigskip

\textbf{Authors:} 
\cc{Note: first 3 authors must be from ESO countries}
Mark Booth (UK Astronomy Technology Centre, UK);
Patricia Luppe (Trinity College Dublin, Ireland);
Sebastian Marino (University of Exeter, UK);
Joshua B. Lovell (Center for Astrophysics, Harvard \& Smithsonian, USA);
Jonathan P. Marshall (Academia Sinica Institute of Astronomy and Astrophysics, Taiwan);
Gaspard Duch\^ene (UC Berkeley, USA / University of Grenoble Alpes, France);
Isabel Rebollido (European Space Astronomy Centre, Spain);
Mark C. Wyatt (University of Cambridge, UK);
Riouhei Nakatani (University of Milan, Italy);
Aya E. Higuchi (Musashino University, Japan);
Miguel Chavez-Dagostino (Instituto Nacional de Astrof\'isica, \'Optica y Electr\'onica, Mexico);
Hiroshi Kobayashi (Nagoya University, Japan)
\linebreak

\textbf{Endorsers}:
Jean-Charles Augereau (University of Grenoble Alpes, France); Raphael Bendahan-West (University of Exeter, UK); Jane Greaves (Cardiff University, UK); Attila Mo\'or (Konkoly Observatory, Hungary); David J. Wilner (Center for Astrophysics, Harvard \& Smithsonian, USA)
\linebreak

\textbf{Science Keywords:} 
exoplanets: planet system architecture; planet-disk interactions; stars: circumstellar matter
\linebreak

\vspace{3cm}

 \captionsetup{labelformat=empty}
\begin{figure}[h]
   \centering
\includegraphics[width=.9\textwidth]{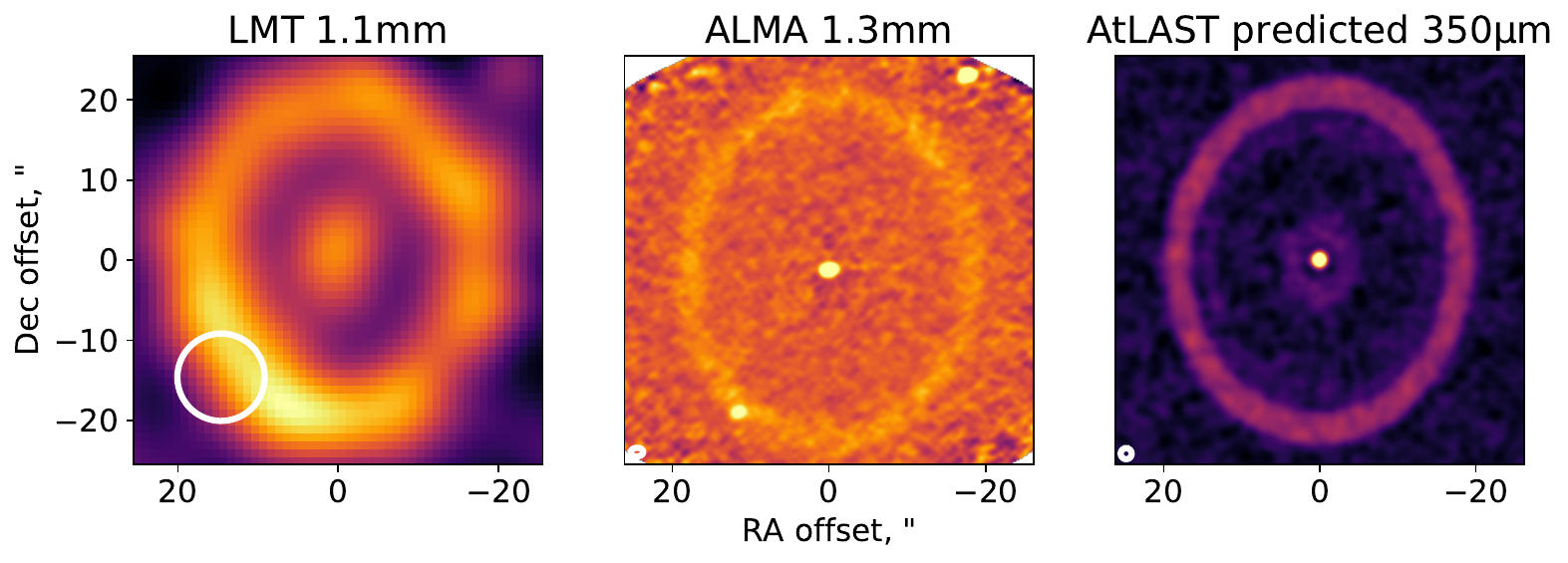}
   \caption{}
\end{figure}
\vspace{-15mm}
}
\setcounter{figure}{0}
\captionsetup{labelformat=default}

\thispagestyle{empty}
\pagebreak
\setcounter{page}{1}

\cc{Below are some suggested section headings for the main 3-pages long document. Modify/delete/disregard as needed to best suit your white paper.} \cc{A reminder that the request from ESO is that the white papers should include:}
\cc{\begin{itemize}
    \item a focus on one (or group of) science question(s) and explain why this needs a facility we do not expect to have by the 2030s
    \item a short ($<$ half page) description of what technology developments / data handling requirements may be needed.  %
\end{itemize}}

\section*{Abstract}

\cc{Suggest to keep it short, 4-5 lines}

Debris discs reveal the architectures and dynamical histories of planetary systems. Sub-millimetre observations trace large dust grains within debris discs, revealing their bulk properties. Debris discs have so far only been detected around $\sim$20\% of stars, representing the bright end of the population. A new facility is required to reach fainter discs, overcoming the confusion limit, with multiwavelength capabilities for characterisation, sensitivity to large-scale emission for nearby targets and a large field of view for surveying distant populations. All of this is made possible with the Atacama Large Aperture Submillimetre Telescope (AtLAST).

\section{Scientific context and motivation}

\cc{Target audience: expert astronomers, but not necessarily in your field. Could be good to keep the motivation to a similar level as the background section of an ALMA large proposal.}

\cc{Suggestions for content:}
\cc{\begin{itemize}
    \item Focus on one (or a group of) science question(s) of interest
    \item 1-2 catchy self-explanatory Figures/graphics with captions to help explain your points.
    \item Explain why this science case cannot be done without AtLAST (why it cannot be done with any other current or planned facility?)
\end{itemize}}

One of the greatest successes of the last century was the detection of planetary systems around other stars via the discovery not only of planets but also of circumstellar discs. The planet formation process leaves behind significant amounts of leftover material in the form of planetesimals and dust, which can remain in the system for Gyr as they are sustained through collisions. The planetesimals tend to be concentrated in belts, like our own Kuiper and asteroid belts, whilst the dust can be distributed throughout the system and so this collection of planetesimals and dust is referred to as a debris disc. They have so far been discovered around $\sim20\%$ of nearby stars (see e.g. Matthews et al. 2014). However, these are much brighter than the Kuiper belt and asteroid belt. Therefore, to understand the context of the Solar System, we need to discover and characterise the population of faint debris discs that are currently below our detection limits.

The first debris discs were detected with IRAS, identifying infrared excesses for a handful of objects (Aumann et al. 1984, Backman \& Paresce 1993). 
The first resolved image came via a scattered light observation of $\beta$ Pictoris (Smith \& Terrile 1984). Further resolved images were not made until the late 1990s, in part thanks to the installation of SCUBA at the JCMT (Holland et al. 1998). This demonstrated the power of submillimetre observations for detecting the thermal emission from cold dust in planetary systems that would otherwise need to be detected from space. Despite starting operations in 1987, the JCMT remains the largest submillimetre single-dish telescope. At 15m, it is capable of only following up the brightest discs detected by space telescopes. Submillimetre interferometers, such as SMA and ALMA, have proved vital to detailed characterisation of these brightest discs, yet prior detections are still needed to ensure efficient planning of these follow-up observations. In order to discover and characterise the faint population of debris discs, a new, large, single-dish, submillimetre telescope is essential. In this white paper, we describe how this can be achieved with the Atacama Large Aperture Submillimetre Telescope (AtLAST). AtLAST is a proposal for a 50~m diameter, single dish with a maximum instantaneous field of view of 2$^{\circ}$ diameter, located at a 5000~m altitude site near ALMA and operating from 0.3-10~mm (Mroczokowski et al. 2025).

This paper summarises results presented in Klaassen et al. (2024), which demonstrated the value of AtLAST for detecting
\begin{itemize}
    \item mature debris discs as faint as our own Kuiper belt at high sensitivity and with a resolution of a few arcseconds, whilst retaining sensitivity to structures of tens of arcseconds;
    \item the epoch of planet and planetesimal formation through large surveys of young stellar objects (YSOs) in star forming regions, the efficiency of such surveys substantially benefiting from the combination of a high sensitivity and large field of view, and the higher frequencies in the sub-millimetre domain. 
\end{itemize}

\vspace{-3mm}
\section{Science case}

\subsection{Detecting faint, nearby debris discs}

\begin{wrapfigure}{R}{0.5\textwidth}
   \centering
   \vspace{-5mm}
\includegraphics[width=.5\textwidth]{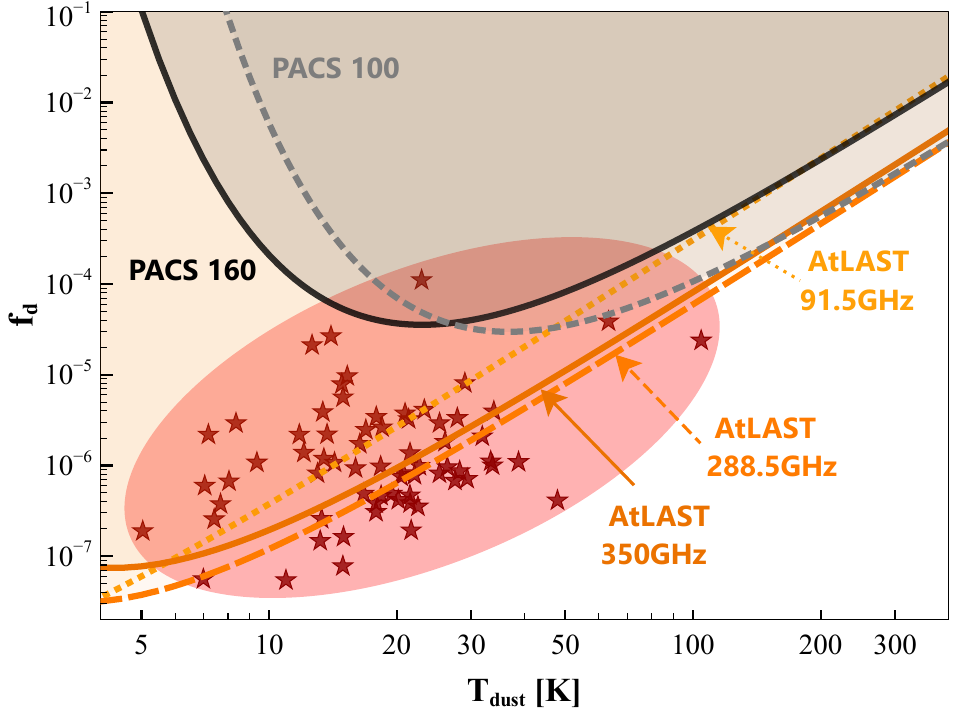}
   \caption{Fractional luminosity as a function of temperature for a simulated population of debris discs around nearby M dwarfs. The lines show approximate detection limits for Herschel/PACS and AtLAST. %
   }
   \label{flum}
\end{wrapfigure}

Studying debris discs around the nearest stars offers a unique window into the composition, architecture and evolution of planetary systems in our cosmic neighbourhood. Through radial velocity, direct imaging and astrometry observations, we are building up a picture of the nearest exoplanets. 
With upcoming facilities such as the Extremely Large Telescope (ELT), the Habitable Worlds Observatory (HWO), and the Large Interferometer For Exoplanets (LIFE) mission poised to deliver unprecedented sensitivity and resolution, we are entering an era where direct imaging and detailed characterization of both discs and planets will become routine. Complementary observations in the sub-mm will allow us to link disc properties to planetary architectures, shedding light on the conditions that foster habitable worlds.

The Kuiper belt is much fainter than other known debris discs and is estimated to have a fractional luminosity\footnote{This is the ratio between the bolometric luminosity of the disc and that of the star} of $5\times10^{-7}$ (Poppe et al. 2019), whereas the faintest debris discs that have been detected so far are typically at least an order of magnitude brighter (Montesinos et al. 2016; Sibthorpe et al. 2018). These detections primarily come from far-IR observations. The discs so far detected at sub-mm wavelengths are at least two orders of magnitude brighter (Holland et al. 2017). %
 The main limitation on reaching such low fractional luminosities with a sub-millimetre single-dish telescope is the extragalactic confusion limit. At sub-mm wavelengths, a 50-m dish is necessary to ensure that the confusion limit is low enough to reach Kuiper belt levels of dust. %
Whilst interferometers can also beat the confusion limit, they also filter out the large-scale emission (on the order of arcseconds for ALMA at the shortest wavelengths) resulting in a lack of understanding in the extended emission of nearby discs that can be many 10s of arcseconds in extent. A single-dish telescope ensures capturing the total flux (and thereby the total mass) of debris discs around our closest neighbours.

The advances in detection rates that will be possible with AtLAST make an unbiased survey of the nearest stars in the submillimetre viable for the first time. 
In terms of detecting faint discs, 860~$\mu$m observations will allow us to reach Kuiper belt levels of dust around Sun-like stars within 10~pc within an hour of observing time per star (going down to a sensitivity of 21~$\mu$Jy\,beam$^{-1}$ -- 3 times the confusion noise). 
For M dwarfs, this is expected to increase the number of detected discs by an order of magnitude (see figure \ref{flum} and Luppe, in prep.), providing us with a statistically meaningful sample of these discs that have, so far, largely avoided detection. 
Additional observations at 350~$\mu$m are necessary to reach spatial scales as small as 1.7 arcseconds (a few au around the nearest stars), thereby resolving all discs as small as the Kuiper belt around stars within about 20~pc. Not only will this help in understanding the structure of the discs, but it will also help avoid background confusion. 
Combining multi-wavelength observations also provides a measure of the spectral slope necessary to our understanding of collisional processes within debris discs. 
This survey will greatly increase our knowledge of planetary systems in the Solar neighbourhood and provide us with the statistics necessary to determine how disc properties correlate with stellar properties. 

\subsection{Detecting the birth of debris discs}
Young Stellar Objects (YSOs) form in star-forming regions. They begin life surrounded by protoplanetary discs that eventually evolve into planets and debris discs. By which point the YSOs are referred to as class III. Class III YSOs offer a view into the early stages of development of young planetary systems. By studying them in the sub-mm, we can learn about their masses, morphologies, grain-size distributions and dispersal mechanisms at this key evolutionary stage (Lovell et al. 2021). Despite their importance to planetary system evolution, class III stars remain severely understudied at sub-mm wavelengths, primarily due to a lack of sub-mm observatories with a sufficient mapping speed to observe 100s-1000s of objects with a sensitivity to the faint emission from their typical dust levels. The rapid decline of dust mass suggested by Lovell et al. (2021) is still not well established, and to address this, an unbiased survey with AtLAST is essential, as it could provide the first decisive constraints on this.

With a 2$^{\circ}$ FoV (filling 10s of thousands of ALMA primary beam areas), and simultaneous deep, multi--band continuum coverage, AtLAST offers a unique opportunity to study this aspect of planetary system evolution. Within a few thousand hours, it will be capable of mapping all the star-forming regions within 1~kpc, drastically increasing the number of known class III YSOs by at least an order of magnitude. A sensitivity limit of, for example, $\sim30\,\mu$Jy will enable the detection of discs down to 0.02~M$_\oplus$ and thereby allow the discovery of a population of young debris discs out to several hundred parsecs, that currently we only have systematically studied in the Sun's neighbourhood, at older ages.

\section{Technical requirements}
\cc{The guidelines recommend to keep this section to less than 0.5 page, describing what technology or data handling requirements may be needed to do the science. The ESO call explicitly notes they are not looking for a detailed description of a facility.}

The key technical requirements for the study of debris discs near and far are:
\begin{itemize}
    \item Observations at sub-mm wavelengths (ideally between 350 and 870~$\mu$m -- i.e. wavelengths short enough to be close to the peak emission and long enough that dust detected traces the parent planetesimal belts.
    \item The ability to detect emission on scales up to an arcminute to ensure we gain a complete picture of the dust distribution around the nearest stars. I.e. a single-dish is required to avoid the flux losses suffered by interferometers.
    \item An angular resolution of $<5''$ at wavelengths $<1$~mm (i.e. a 50~m primary mirror). This ensures the confusion limit is low enough at sub-mm wavelengths to be able to reach dust levels as low as the Kuiper belt around nearby stars.
    \item In order to rapidly map nearby star-forming regions, a large field of view of at least $1^\circ$ is required.
\end{itemize}

AtLAST satisfies all the requirements of this science case along with a wide range of other science cases, as summarised in Booth et al. (2024).

\section*{References}
{
Aumann H. H., et al. 1984, ApJL, 278, L23-L27 
$\bullet$ Backman D. E., Paresce F. 1993, prpl.conf, 1253 
$\bullet$ Booth M., et al. 2024, arXiv:2407.01413 
$\bullet$ Holland W. S., et al. 1998, Natur, 392, 788-791 
$\bullet$ Holland W. S., et al. 2017, MNRAS, 470, 3606-3663
$\bullet$ Klaassen P., et al. 2024, ORE, 4, 112
$\bullet$ Lovell J. B., et al. 2021, MNRAS, 500, 4878-4900 
$\bullet$ Matthews B. C., et al. 2014, prpl.conf, 521-544 
$\bullet$ Montesinos B., et al. 2016, A\&A, 593, A51 
$\bullet$ Mroczkowski T., et al. 2025, A\&A, 694, A142 
$\bullet$ Poppe A. R., et al. 2019, ApJL, 881, L12
$\bullet$ Sibthorpe B., et al. 2018, MNRAS, 475, 3046-3064 
$\bullet$ Smith B. A., Terrile R. J. 1984, Sci, 226, 1421-1424 
}

\end{document}